\documentclass[doublespacing]{elsart}
\usepackage{epsfig}

\begin{document}

\begin{frontmatter}
\title{An Off-lattice Model for Br Electrodeposition on Au(100):
from DFT to Experiment}

\author{S.J.~Mitchell and M.T.M.~Koper}

\address{
Schuit Institute for Catalysis and Department of Chemical Engineering,\\
Eindhoven University of Technology,\\
5600 MB Eindhoven,\\
The Netherlands
}

\begin{abstract}
Since Br adsorption on Au(100) displays an incommensurate ordered phase,
a lattice-gas treatment of the adlayer configurations is not reliable.
We therefore use density functional theory slab calculations 
to determine the parameters necessary for the construction of an off-lattice model.
We compute and analyze the total energy and electron density as the lateral Br position and coverage are varied.
This allows the calculation of the corrugation potential, the short-range lateral interactions,
the dipole moment (long-range interactions), and the residence charge.
From these parameters,
we construct an off-lattice model with no freely adjustable parameters.
The simulation results compare remarkably well with experimental results.
\end{abstract}


\end{frontmatter}

\section{Introduction}
\label{intro}
In recent years,
a variety of theoretical studies of electrochemical adlayer formation have emerged 
as a result of new {\it in situ} experimental techniques,
which provide atomic-scale structural information about the adlayer.
For a review of Monte Carlo studies, see Ref.~\cite{ME:ECCHAPT}.
Among these studies, halide adsorption, in particular, has provided a wealth of fundamental insight into the structure
and formation of electrochemical adlayers \cite{MAGNUSSENREV}.
Bromine adsorption on Ag(100) stands out as one of the most extensively studied examples,
both experimentally and theoretically
\cite{ENDO99,IGNACZAK:QMHALIDE,KOPER:HALIDE,MITC00A,MITC00C,MITC00B,OCKO97,MITCH01A,MITC02,OCKO:BR/AG,VALETTE:AG,WANG:FRUM}

The Br/Ag(100) system displays a second order phase transition from a disordered adlayer at low potentials
to a $c(2 \times 2)$ adlayer at higher potentials.
The most successful theoretical treatments of the Br/Ag(100) adlayer use a lattice-gas approximation
\cite{KOPER:HALIDE,MITC00A,MITC00C},
in which the adsorbates are assumed to occupy an array or lattice of adsorption sites.
However, when a system displays incommensurate ordered phases,
the lattice-gas approximation is not a valid assumption,
since the periodicity of an incommensurate phase and the periodicity of the substrate surface do not match.
In these cases, an off-lattice model must be constructed,
which consists of laterally interacting adsorbates in a two-dimensional corrugation potential.
A full and formal description of an off-lattice model for Br/Ag(100) can be found in Ref.~\cite{MITC02},
and the lattice-gas approximation is seen to be a special case of the off-lattice model in which the
amplitude of the corrugation potential goes to infinity.

For a more complete understanding of simple adsorption phenomena,
we should also consider systems which display incommensurate phases,
and bromide adsorption on Au(100) is a good candidate for studying commensurate to incommensurate phase transitions.
{\it In situ} Surface X-ray Scattering (SXS) \cite{WAND96} and Scanning Tunneling Microscopy (STM) \cite{KOLB00}
measurements of Br adsorbed on Au(100) in solution, 
as well as {\it ex situ} Low Energy Electron Diffraction (LEED) measurements \cite{RENIERS99},
confirm that Br adsorbs preferentially at the bridge site of the Au(100) substrate.
Three phases are observed for Br on Au(100):
a disordered phase at low coverages,
a commensurate ordered phase with coverage $\Theta=1/2$,
and a uni-axially compressed incommensurate phase with $\Theta>1/2$.
Two points of theoretical interest are immediately obvious.
First, since the system displays the same binding site and ordered phases in both vacuum and in solution,
the need to include water into the theoretical model is greatly reduced,
as water does not seem to play an active role in the adlayer formation.
Second, the presence of an incommensurate phase excludes a lattice-gas treatment of the adlayer,
in contrast to the Br/Ag(100) system.

In previous theoretical studies, 
symmetry arguments and parameter fitting have been the primary methods used for deriving the lattice-gas model parameters.
These techniques have worked well for systems requiring only a single parameter,
like Br/Ag(100) \cite{KOPER:HALIDE},
but these techniques for deriving model parameters have met with varied success for more complex systems,
like Cu and sulfate co-adsorption \cite{BROW99}.
With the introduction of the corrugation potential, necessary for an off-lattice model,
the complexity of the parameter space is greatly increased,
and therefore, in this paper, we use density functional theory (DFT) to estimate the off-lattice model parameters for Br/Au(100).

Before discussing the details of the DFT slab calculations,
it is important to point out several issues which may be of physical importance 
but which were not or could not be included in the present work. 
First, there is no effect of an electrode potential in the DFT calculation methods used here,
either as an external electric field 
or, as would be preferred, in the form of charging and discharging of the substrate.
This effect is not included in the calculations because it is difficult and time consuming,
and there is not yet a generally accepted method for its inclusion \cite{KOPER03}.
DFT cluster calculations of a single halide adsorbate in the presence of an external electric field have, 
however, been performed by other authors \cite{WASILESKI02}, 
but since macroscopic scale experimental systems contain many adsorbates,
we expect the coverage to also represent an important degree of freedom.

The second issue of possible physical importance is the inclusion of liquid water into the supercell. 
This is currently possible using a hybrid DFT/Monte Carlo or DFT/molecular dynamics 
technique to perform proper thermal sampling, 
but this is extremely computationally intensive. 
Despite the lack of water in the supercell, 
the model parameters derived here seem quite reasonable,
as justified by the results of the off-lattice Monte Carlo simulations,
presented in Section~\ref{sec:offlatt}.

This paper is organized as follows.
Section~\ref{sec:methods} discusses the details of the DFT calculations and the analysis methods used to interpret the DFT results.
Section~\ref{sec:resdisc} presents the off-lattice model parameters and discusses their reasonableness and physical significance.
Section~\ref{sec:offlatt} presents the results of an off-lattice model for Br/Au(100)
and compares these results to experiment.
Finally, Sec.~\ref{sec:conc} presents a summary and conclusions.

\section{Methods} \label{sec:methods}

All DFT results presented here were calculated using the Vienna Ab Initio Simulation Package (VASP) \cite{VASP1,VASP2,VASP3} 
using a plane-wave basis set, 
a periodic slab geometry, 
the generalized gradient-corrected exchange-correlation functional (GGA, PW91) \cite{GGA}, 
Vanderbilt ultra-soft pseudopotentials \cite{VANDERBILT90}, 
and a cut-off energy of 400~eV. 
All slabs contained three metal layers separated by six vacuum layers, 
each vacuum layer being the equivalent thickness of a metal layer. 
To examine possible coverage dependence, supercells of two different sizes were examined, 
a $2 ( 2.97 ) \times 2 ( 2.97 ) \times 12.6$~\AA{} cell (denoted as $2 \times 2$),
containing 4 surface Au atoms, 
and a $3 ( 2.97) \times 3 ( 2.97 ) \times 12.6$~\AA{} cell (denoted as $3 \times 3$),
containing 9 surface Au atoms.
The $k$-point mesh was generated using the Monkhorst method with a $7 \times 7 \times 1$ grid for the $2 \times 2$ cells 
and a $5 \times 5 \times 1$ grid for the $3 \times 3$ cells. 
Initial convergence tests indicated that these parameters are sufficient for the accuracy desired,
as justified by the results in Sec.~\ref{sec:offlatt}.

The Au(100) surface normal was oriented in the $+z$ direction. 
A typical DFT supercell is shown in Fig.~1.
Unless otherwise indicated, all Au(100) slab atoms were located at their bulk positions with a
face-centered-cubic lattice constant of 4.2~\AA{}, 
as determined by our own {\it ab initio} calculations. 
This yields a (100) surface lattice constant of $a = 2.97$~\AA{}, 
as compared with the measured value of 2.885~\AA{} \cite{WAND96}, an error of 3\%.

In many of the most common applications of DFT methods, 
the positions of all or some of the atoms in the configuration are relaxed 
in such a way as to minimize the total energy. 
However, since we are interested in more quantities than just the total energy 
and we wish to compare results for different coverages, 
a full relaxation of the initial atomic configuration is not desirable for two reasons. 
First, even though Au(100) is known to display a surface reconstruction, 
the surface reconstruction is lifted by even moderate halide coverages \cite{WAND96}, 
thus eliminating the need to reproduce such a transition in the calculations 
by relaxation of the slab in the surface tangent directions. 
Second, early in this research, 
an examination of the charge density revealed very large surface dipole moments 
for buckled surface configurations,
e.g.\ when the surface atoms were relaxed independently in the surface normal direction.
These large surface dipole moments made analysis and understanding of the Br-Au(100) bond polarity
extremely difficult and problematic.
Therefore, when any relaxations were performed, 
they were performed by keeping all slab atoms at their bulk positions, 
except for the top layer atoms (Br-containing surface layer) which were relaxed simultaneously in a plane,
and the Br layer was relaxed in a similar way. 
We introduce the notations $z_{\rm top}$ and $z_{\rm Br}$
which indicate the $z$ positions of the top slab layer and the Br adsorbate layer, respectively.
As is well known, 
relaxation of the top metal layers in slab calculations generally has a relatively small effect on the adsorption energy,
the effect being of the order of 5 to 100~meV,
which is in agreement with the energy differences observed in this work between the fully relaxed and unrelaxed substrate.

In addition to obtaining total energies, 
DFT results also include the electronic charge density, $\rho_{\rm e } ( \vec{x} )$, 
which has the sign convention of being more positive where the density of electrons is higher.
Figure~2 shows the electronic charge density,
$\rho_{\rm e} ( z ) = \int \rho_{\rm e} ( \vec{x} ) \; dx dy$, 
for several binding site configurations. 
The location of the three metal layers and the Br adlayer are clearly visible, 
and outside of the surface region, 
few differences can be seen between the different adsorbed states.

\subsection{Charge Transfer Function}

Unfortunately, the electron density, $\rho_{\rm e} ( \vec{x} )$ or $\rho_{\rm e } ( z )$, 
when plotted in its unmodified form (see Fig.~2), 
is difficult to interpret, 
and little insight is gained from so simple a plot. 
Even the full charge density, $\rho ( \vec{x} )$, including both electrons and ionic cores,
gives little novel understanding,
and only when the ``background'' charge is eliminated, 
is the presence of the polar surface bond easily seen. 
We therefore introduce the charge transfer function, 
so called because it clearly shows where charge has been transferred 
as a result of the formation of the (polar) surface bond. 
The charge transfer function is defined as follows:
\begin{equation}
  \begin{array}{ccc}
    \Delta \rho(\vec{x}) & = & \left\{ \rho(\vec{x})_{N \rm Br-Au(100)} - \sum_i \rho_i (\vec{x})_{\rm Br} - \rho(\vec{x})_{\rm Au(100)} \right\} / N\\
    & = & - \left\{ \rho_{\rm e} (\vec{x})_{N \rm Br-Au(100)} - \sum_i \rho_{{\rm e}i} (\vec{x})_{\rm Br} - \rho_{\rm e} (\vec{x})_{\rm Au(100)} \right\} / N \; ,
  \end{array} \label{eq:chargetransfer}
\end{equation}
where $\rho(\vec{x})_{N \rm Br-Au(100)}$ is the full charge density of the adlayer system with $N$ adsorbed Br in the cell, 
$\rho_i(\vec{x})_{\rm Br}$ is the full charge density of a single Br atom at the same position as that in the Br-Au bonded system, 
$i$ indexes the $N$ adsorbed Br, 
$\rho(\vec{x})_{\rm Au(100)}$ is the charge density of the Au(100) slab with all atoms at the same positions as in the Br-Au bonded system, 
and the subscripts ``e'' denote electron only densities, 
having the sign convention that positive indicates greater electron density. 
The charge transfer function is calculated using only charge densities for the same size cells.

Since all atoms are located at the same positions for each of the charge densities in Eq.~\ref{eq:chargetransfer}, 
the ionic core charges cancel, leading to the second, equivalent expression in Eq.~\ref{eq:chargetransfer}. 
Also, since we here consider only charge neutral cells, the charge transfer function must integrate to zero.
Figure~3 shows $\Delta \rho(z)$ for the same configurations shown in Fig.~2,
and the charge polarization at the surface is easily seen.

\subsection{Residence Charge}

The net charge on the adsorbate is one of the simplest results to calculate from the charge transfer function. 
Unfortunately, there is no unique scheme to assign charge to each atom; 
however, in the simplest method, 
charge above the polarization center is assigned to the adsorbate, 
and the charge transfer function is integrated over this region. 
We define the residence charge as
\begin{equation}
  q = \int_{5.4~\AA{}}^{9.4~\AA{}} \Delta \rho ( z ) \; dz,
\end{equation}
where the limits of integration are chosen between the polarization center and the chargeless region of the vacuum. 
Since the charge transfer function must integrate to zero for charge neutral systems, 
the charge residing on the slab must be equal and opposite to $q$.

\subsection{Surface Dipole Moment}

Motivated by the success of the lateral $1/r^3$ dipole-dipole repulsion 
in describing the ordering behavior of Br electrodeposited onto Ag(100) \cite{MITC00C}, 
we calculate the dipole moment by
\begin{equation}
  \vec{p} = \int ( \vec{x} - \vec{Z} ) \rho ( \vec{x} - \vec{Z} ) \; d x d y d z \; , 
\label{eq:dipmom0}
\end{equation}
where $\vec{Z}$ is the center of the dipole, 
and the integral extends over the entire range of the charge distribution.

From Eq.~\ref{eq:dipmom0}, it would appear that the dipole moment is dependent on the seemingly arbitrary center of the dipole; 
however, we note that when the charge density goes to zero at the limits of integration, 
$\vec{p}$ is independent of $\vec{Z}$ \cite{JACKSON}. 
Since all charge distributions considered in this paper go to zero in the vacuum region, 
we choose the limits of integration in the $z$ direction to be $[ - 3.2, 9.4 ]$~\AA{}, 
and the integration in the surface tangent directions obviously extend over the full $x$ and $y$ ranges of the cell.

From symmetry considerations, the dipole moment must be of the form $\vec{p} =p \hat{z}$, 
at least for the symmetric $(x,y)$ positions (bridge, hollow, and top sites), 
and we therefore introduce the surface dipole moment and define it as follows:
\begin{equation}
\begin{array}{lll}
  s & = & \vec{p} \cdot \hat{z} = \int z \rho ( z )_{N \rm Br-Au(100)} / N \; d z \\
    & = & \int z \left\{ N \Delta \rho(z) + \rho(z)_{\rm Au(100)} + \sum_i \rho_i(z)_{\rm Br} \right\}/ N \; dz \\
    & = & \int z \Delta \rho(z) \; dz + \int z \rho(z)_{\rm Au(100)} / N \; dz + \sum_i \int z \rho_i(z)_{\rm Br} / N \; dz\\
    & = & \int z \Delta \rho(z) \; d z \; ,
\end{array}
\label{eq:dipmom}
\end{equation}
where the factor of $1 / N$ is included to correct for multiple Br-Au(100) dipoles in the cell.
Of course, the charge distribution due to a single isolated Br has no
intrinsic dipole moment and likewise for a symmetric isolated slab. 
Since these terms cannot contribute to the dipole moment, 
we are left with the final, simplified form of Eq.~\ref{eq:dipmom},
and the sign convention is such that the dipole points from positive charge to negative charge.

\section{DFT Results and Discussion} \label{sec:resdisc}

In this section we present results from analyzing both the total energy and the charge transfer function. 
We examine both the lateral dependence and the coverage dependence using a variety of configurations. 
For the lateral dependence studies, 
the lateral and vertical position of a single Br atom in a $3 \times 3$ cell were varied, 
and for simplicity, the slab atoms were fixed at their bulk positions.
From the lateral dependence of the total energy, we can construct a sinusoidal approximation of the surface binding energy, 
which we call the corrugation potential.
From the charge distribution, 
we can calculate the lateral dependence of the surface dipole moment and the residence charge.

To examine the coverage dependence of the total energy and the charge distribution, 
only adlayer configurations with four-fold symmetry were considered, 
and all Br atoms were placed at the preferred bridge binding sites, 
see Fig.~4. 
In addition to the four cells shown in Fig.~4, 
two additional $2 \times 2$ cells were examined with four and eight adsorbed Br, 
corresponding to coverages of $\Theta = 1$ and $\Theta = 2$, respectively. 
Although these highest two coverages are expected to be unphysical, 
their calculation was necessary to determine the short range excluded volume repulsion, 
needed for an off-lattice Monte Carlo model.

\subsection{Corrugation Potential}

We define the corrugation potential as the total energy of the adsorbate system, 
minimized with respect to $z_{\rm Br}$, as a function of $x$ and $y$. 
Experiments in both the {\it in situ} and {\it ex situ} environments \cite{WAND96,RENIERS99} 
confirm that Br preferentially binds at the bridge site, 
indicating that the corrugation potential is lowest at this site; 
however, this knowledge alone is not sufficient to determine the symmetry or amplitude of the corrugation potential. 
DFT calculations must therefore be used to determine the corrugation potential.

For each $x, y$ location (see Table~\ref{tab:min}) $z_{\rm Br}$ was varied systematically. 
Figure~5 shows the total energy as a function of $z_{\rm Br}$ for several important $x, y$ locations. 
The total energy minimum for each $(x,y)$ position, 
along with the $z$ value of the minimum are reported in Table~\ref{tab:min}. 
For clarity, the total energy of the bridge site minimum has been subtracted from each, 
and the Br sits just above the third metal layer which is centered at $z_{\rm top} = 4.2$~\AA{}.

A table of discrete results gives little impression of the surface symmetry,
and since the off-lattice Monte Carlo simulations (see Sec.~\ref{sec:offlatt})
require the corrugation potential to be defined at all $(x,y)$ points,
we construct a sinusoidal approximation to the DFT results of Table~\ref{tab:min}.
Certainly, many different approximations can be made; 
however, we favor the following simple form which reproduces the DFT results extremely well around the bridge site, 
the region of most importance.
\begin{equation}
  \begin{array}{cc}
    U ( x, y ) = & \Delta_1 \left[ \cos \left( \frac{2 \pi x}{a} \right) \cos
    \left( \frac{2 \pi y}{a} \right) + 1 \right] / 2\\
    & + \Delta_2 \left[ \cos \left( \frac{2 \pi x}{a} \right) + \cos \left(
    \frac{2 \pi y}{a} \right) \right] / 2 \hspace{0.75em},
  \end{array} \label{eq:U}
\end{equation}
where $\Delta_1 = 274$~meV and $\Delta_2 = 234$~meV.

For comparison with the original DFT results, 
values of the sinusoidal approximation are also listed in Table~\ref{tab:min}, 
and the full sinusoidal approximation is shown in Fig.~6.
The differences between the DFT calculations and the sinusoidal approximation are expected 
to have little qualitative effect on the equilibrium behavior of the system, 
since these differences appear primarily at the low-probability binding positions. 
We also comment that the symmetry and amplitude of $U(x,y)$ are consistent 
with the observed ordered phases and phase transitions \cite{WAND96},
as will be made clear in Sec.~\ref{sec:offlatt}.

\subsection{Lateral Interactions}

Assuming only two-body interactions, 
the effective lateral interactions between adsorbed Br can be determined 
by analyzing the total energy for configurations representing a range of coverages. 
Two things must be noted.
First, to simplify the analysis, $z_{\rm top} =$4.2~\AA{} and $z_{\rm Br} =$6.4~\AA{} in all calculations. 
Second, since we wish to compare results at different coverages, 
it is imperative that each adlayer should have the same square symmetry, 
thus changing the coverage has only the effect of changing the Br-Br nearest-neighbor separation, $a_{\rm Br}$,
as illustrated in Fig.~4.

To separate contributions to the total energy from contributions other than lateral interactions, 
we assume that the total energy for each adlayer configuration can be approximated as
\begin{equation}
E_{\rm tot} = E_{\rm slab} + N E_{\rm Br} + N \mathcal{B} + \Phi \; , 
\label{eq:normphi}
\end{equation}
where $E_{\rm slab}$ is the total energy of the blank Au(100) slab,
$E_{\rm Br}$ is the energy of a single Br atom, 
$N$ is the number of Br in the supercell, 
$\mathcal{B}$ is the binding energy of a single adsorbate to the substrate, 
and $\Phi$ is the total lateral interaction energy between adsorbates. 
The values of the terms $E_{\rm slab}$ and $E_{\rm Br}$ can be determined from calculations including 
only the Au(100) slab and only a single Br atom, respectively.

The results of this calculation are shown in Fig.~7
and include results for the four coverages shown in Fig.~4 plus two additional cells with coverages $\Theta = 1$ and 2,
corresponding to every other bridge site occupied and every bridge site occupied, respectively. 
These later two systems were included to examine the short range repulsion, 
which is a crucial parameter needed to construct an off-lattice Monte Carlo model. 
A non-linear fit to the data shown in the figure indicates
\begin{equation}
  \Phi / N + \mathcal{B} = 7 \times 10^5 a_{\rm Br}^{- 6} - 2560 \; ,
\label{eq:short}
\end{equation}
with units of meV,
and thus, $\mathcal{B}=-2560$~meV.
Note that the fit is primarily determined by the first three points,
where the repulsion is largest,
and this indicates that the $r^{-6}$ repulsion represents the effective short-range ionic core repulsion.

\subsection{Surface Dipole Moment}

The polar surface bond can be measured by integrating over $x$ and $y$ to obtain $\Delta \rho(z)$, 
shown for various binding sites in Fig.~3. 
All sites show a clear polarization of charge in the surface region and an oscillatory charge in the first two slab layers;
however, the relative magnitude of these two behaviors is quite different 
between the four sites and is quite sensitive to site coordination.
The quarter and bridge sites are seen to be qualitatively similar, 
both displaying a pronounced charge polarization in the surface region with moderate oscillations in the slab surface region. 
The top and hollow sites, however, are quite different from each other and from the quarter and bridge sites, 
the top site having the largest oscillations in the slab surface region and the hollow site having the smallest oscillations.

Figure~9 shows the surface dipole moments for all considered configurations in the lateral dependence study. 
There is very little direct dependence on the lateral position, 
but there is a strong, almost linear, dependence on $z_{\rm Br}$.
The heavy line in Fig.~9 indicates a linear fit to all $s$ values and has a slope of 1.22~e, 
where e is the elementary charge unit. 
This suggests that Br/Au(100) forms an ionic bond, 
regardless of the coordination with the substrate.
A two-dimensional slice through $\Delta \rho$, Fig.~8,
confirms the p-type character of the electron-accepting orbital on Br.

Although there is no direct dependence of $s$ on the $(x,y)$ position, 
the total energy does have a strong dependence on $(x,y)$, 
and this leads to an indirect lateral dependence of the surface dipole moment, 
when only the most energetically favorable $z_{\rm Br}$ are considered. 
These values are summarized in Table~\ref{tab:min},
and from these values, we can construct an approximate sinusoidal form needed for off-lattice Monte Carlo simulations,
\begin{equation}
  s(x,y) = 0.08 \left\{ h(y)g(x) + g(y)h(x) \right\} + 0.14 \; , 
\label{eq:sxy}
\end{equation}
where $h ( x ) = \cos ( 2 \pi x / a ) + 2$, $g ( x ) = \cos ( 2 \pi x / a ) + 1$, and the units are e\AA{}.

For comparison with the DFT results, the values of $s ( x, y )$ at several lateral positions are also listed in Table~\ref{tab:min}, 
and the sinusoidal approximation is obviously very good at positions with low values of the corrugation potential, 
where the adsorbate is preferentially located. 
For the top and surrounding sites, 
the approximation given by Eq.~(\ref{eq:sxy}) is quite poor; 
however, since the adsorbate spends very little time at positions where the corrugation potential is large, 
the top region is not expected to significantly influence the behavior of the adlayer.

The surface dipole moment, $s$, can also be calculated as a function of coverage, 
and the results are shown in Fig.~10 for all values of $z_{\rm top}$ and $z_{\rm Br}$. 
The shaded symbols in the figure indicate the configurations corresponding to the energy minimum for each coverage. 
The results are summarized in Table~\ref{tab:cov}.
These results are to be expected from a simple examination of $\Delta \rho(z)$ for the different coverages.
See Fig.~11.
No significant differences are seen between the four curves,
suggesting that there is no significant coverage dependence of either the dipole moment or the residence charge.
This can also be confirmed by the linear dependence of the Fermi energy, $E_{\rm F}$,
on coverage.
See Fig.~12.

\subsection{Residence Charge}

The lateral and coverage dependence of the residence charge are summarized in Tables~\ref{tab:min} and \ref{tab:cov}. 
From these data, we can conclude that $q = - 0.12 \pm 0.01$,
independent of coverage,
with only bridge-site binding considered.
Assuming that the double layer is dominated only by the Helmholtz layer,
the residence charge can be related to the electrosorption valency as $\gamma = z - q/e=-1-q/e$, 
where $z$ is the charge state of the ionic adsorbate species in solution
\footnote{By assuming a parallel plate capacitor of area, $A$, total charge, $q\Theta A$, 
a plate separation, $d$,
and a dipole moment $s=qd$,
Eq.~18.18 of Ref.~\cite{SCHM96} reduces to $\gamma=z-q/e$.
}.
This leads to a rough estimate of the electrosorption valency as $\gamma= - 0.9 \pm 0.1$,
with no direct coverage dependence,
and this estimate still allows for the possibility of a complete, or nearly complete, 
discharge of an adsorbed Br$^-$ ion as suggested in Ref.~\cite{WAND96}. 
However, since these calculations do not include the effect of excess surface charging or a external field,
the connection between the calculated residence charge and the electrosorption valency is still unknown.

A classical charge/image charge interpretation of this system is not correct.
This can be illustrated more completely by examining the behavior of a point charge approximation of $\Delta \rho(z)$,
which can always be constructed,
even when the system is not charge/image charge like.
The point charge approximation consists of two point charges of $q$ and $-q$, 
located at the centers of charge of the adsorbate and the slab, respectively,
and separated by a distance $d$,
such that $s=qd$.
As shown in Fig.~10,
$s$ has a linear dependence on $z_{\rm Br}$ with slope $ds/dz_{\rm Br}=1.22e$.
If the charge/image charge interpretation is correct,
then $q(z_{\rm Br})$ would be a constant, 
and the slope would be $ds/dz_{\rm Br}=2q$.
However, as Fig.~13 shows,
$q(z_{\rm Br})$ is not a constant,
clearly indicating that the system is not charge/image charge like.

\section{Off-lattice Simulations} \label{sec:offlatt}

The off-lattice Monte Carlo simulation method used here is discussed in complete detail in Ref.~\cite{MITC02},
and there is no need to repeat such a detailed discussion here.
The simulated surface was an $L\times L$ square with periodic boundary conditions,
where $L$ was the number of surface atoms in a surface lattice direction.
The corrugation potential used in the off-lattice simulations is given in Eq.~\ref{eq:U},
and the pair-wise lateral interactions between adsorbed Br are the summation of the strong short-range $1/r^6$ repulsions
from Eq.~\ref{eq:short} and a much weaker long-range dipolar repulsion.

There are several different possible treatments for the dipolar repulsion.
One can assume that the dipole moment is flat or constant across the surface,
that is $s(x,y)=s$,
or one can assume the functional form of Eq.~\ref{eq:sxy}.
Also, one can assume a simple $1/r^3$ dipole repulsion,
known as the point dipole approximation, as used in Refs.~\cite{MITC00C} and \cite{MITC02},
or alternately, one could use the more complex physical dipole form,
involving the electrostatic repulsion of four different point charges.
In this paper, we take the simplest assumption for the dipole-dipole repulsion,
namely that of a constant dipole moment, $s(x,y)=s_{\rm bridge}=0.30$~e\AA{},
with the simple $1/r^3$ point dipole form used in Refs.~\cite{MITC00C} and \cite{MITC02}.
The lateral interactions are truncated for $r> 5 a$,
and for $r \le 5 a$,
\begin{equation}
\phi(r \le 5 a)=C_1 s^2 r^{-3}+C_2 r^{-6} \; ,
\end{equation}
where $r$ is the distance between interacting adsorbed Br,
measured in \AA{},
$s=0.3$~e\AA{},
and $C_1=14387$ is the conversion factor which gives $\phi$ in units of meV,
and $C_2=300000$ is the constant which gives $\phi$ in units of meV.

The $\Theta(\bar{\mu})$ isotherm results of the this model are shown in Fig.~14,
along with four insets showing the two-point correlation function for each of the four observed phases.
The two-point correlation function, $C(\Delta r_x, \Delta r_y)$,
is simply the probability of finding an adsorbate at a relative position of $(\Delta r_x, \Delta r_y)$ from any other adsorbate.
Thus, there is always a sharp peak at $C(0,0)$,
which is the self correlation point.
To improve the visualization,
the self-correlation point, $C(0,0)$, has been removed from the insets,
and the grayscale of each inset has been scaled between the highest value of $C(\Delta r_x, \Delta r_y)$ (white)
and the lowest value (black).

The correlation functions in Fig.~14 can be viewed as a kind of average snapshot of the adlayer
and are related to surface X-ray scattering data \cite{MITC00C}.
From left to right, the correlation functions in Fig.~14 represent
a disordered commensurate phase ($\bar{\mu}=600$~meV),
an ordered $c(\sqrt{2} \times 2 \sqrt{2})R45^\circ$ commensurate phase with $\Theta=1/2$ ($\bar{\mu}=1000$~meV),
and two apparently different uni-axially compressed incommensurate phases ($\bar{\mu}=1060$ and 1230~meV),
which represent a distortion between the commensurate ordered phase and an incommensurate hexagonal phase.

The four ordered phases observed in the simulations are quite similar to those found experimentally \cite{WAND96},
except that experiments have observed only one incommensurate ordered phase.
This apparent difference may signify that our model needs some minor adjustments,
such as using a different approximation for the dipole-dipole interactions,
or it may be simply an artifact of the square simulation geometry.
The simulated surface is square with periodic boundary conditions,
but the incommensurate phases are nearly hexagonal.
This difference in geometry obviously creates a strain on the incommensurate adlayer phases,
and this strain may cause slight differences in the observed incommensurate phases.
The strain should decrease with increasing $L$;
however, since these simulations for $L=32$ required about six weeks of computational time 
on a PC with a single GHz processor,
the effect of larger system sizes has not been checked.
As such, the results shown in Fig.~14 should not be considered 
to be the definitive Monte Carlo results of this off-lattice model.

The reader should note that off-lattice simulations are extremely computationally intensive \cite{MITC02},
and commensurate to incommensurate phase transitions depend very sensitively on the specific values of the parameters,
thus making a reproduction of experimental results very difficult.
Our preliminary models for Br/Au(100), based on less accurate Monte Carlo relaxations and sampling,
smaller system sizes ($L=8$ and $L=16$), and less accurate DFT parameter estimations,
usually showed only one incommensurate phase.
However, we prefer to show the more accurate Monte Carlo results of Fig.~14,
even though these may not fit as well to the experiments as the less accurate preliminary results.

Figure~15 shows the experimental cyclic voltamogram (CV) from Ref.~\cite{WAND96}
compared to the simulated quasi-equilibrium CV.
The simulated quasi-equilibrium CV current is
\begin{equation}
i(E)=A \frac{{\rm d}\Theta(\bar{\mu}=B-\gamma e E)}{{\rm d} \bar{\mu}} \; ,
\end{equation}
where $A$ is a constant related to the area of the electrochemical cell and the scanrate,
$\bar{\mu}$ is the electrochemical potential (See Ref.~\cite{MITC02}.),
$B$ is a constant related to the choice of electrochemical reference potential,
$\gamma=-0.9$ is the electrosorption valency as determined from the residence charge,
and $E$ is the experimental electrode potential.

The simulation curve is fit to the experimental curve in Fig.~15 by first calculating 
${\rm d}\Theta/{\rm d} \bar{\mu}$ directly from the Monte Carlo simulation.
Then, the parameters $A$ and $B$ were varied, 
until the simulation and experiment curves matched near the phase transition
from the disordered Br adlayer to the commensurate ordered adlayer.
Since the adjustment of $A$ and $B$ represents simply a scaling of the current and shifting of the potential,
$A$ and $B$ should not be considered as true physical parameters of the off-lattice model,
and as such, the results in Fig.~15 represent a model with no freely adjustable parameters,
since all model parameters were estimated from DFT calculations.

\section{Summary and Conclusions} \label{sec:conc}

Using DFT slab calculations,
we have studied both the lateral and coverage dependence of Br adsorption on Au(100).
We have examined both the total energies and the electron densities,
and their analysis has enabled us to obtain the corrugation potential,
the lateral interactions at short range,
the dipole moment (the interactions at long range),
and the residence charge (the charge remaining on the adsorbate).

We have found that the raw electron densities are difficult to interpret,
and we have therefore defined and analyzed the charge transfer function,
from which we can calculate the surface dipole moment and the residence charge.
Both the surface dipole moment and the residence charge were found to show a strong dependence on the lateral Br position,
but neither showed any noticeable coverage dependence.

It should be noted that the coverage dependence study does not include the effect of the electrode potential or external electric field,
and thus, a full understanding of the coverage dependence, or lack thereof, is not yet known.
We can, however, conclude that the lack of coverage dependence in the electron densities is not
particularly surprising when one considers the energy scales of the adsorption problem.
Electron binding energies are typically of the order of eV, 
whereas the lateral interaction energies at half coverage are of the order of tens of meV.
From these energy considerations, 
we can conclude that simply changing the coverage has little effect on the electron density.

Considering the number of parameters and the sensitivity of off-lattice simulations to the values of the parameters,
we feel that the comparison between simulation and experiment is quite remarkable,
considering that a number of presumably important effects were left out,
such as an external electric field and the presence of liquid water.
We also feel that the similarity of the simulation and experiment justify the values of the DFT parameters,
and despite the differences between the simulation and the experiment in the incommensurate phase,
constructing an off-lattice model from DFT parameters seems to be a valid and predictive method.
Whether this technique will prove valuable for other systems remains to be seen.

Further studies should include the effects of both water and the electrode potential to give further insight into these effects.
However, since Br/Au(100) displays the same ordered phases and binding sites 
in both vacuum and aqueous environments \cite{WAND96,RENIERS99}
and since our simulation results in vacuum compare so well to experiment,
the importance of including water in this DFT study may be less important than in systems which are known to be strongly solvated,
such as with systems containing sulphate.

\section*{Acknowledgments}

This work was funded by the Netherlands Organization for Scientific Research (NWO).
The authors would like to thank T.~Wandlowski for useful discussions, encouragement,
and the use of his experimental data.

%The references should start on their own page.
\clearpage

\clearpage
\begin{table}
\caption{
Summary of results for the Br lateral dependence study.
The nearest Au surface atoms are located at $(x/a,y/a)=$ $(0,0)$, $(1,0)$, $(1,1)$, and $(0,1)$.
The DFT results are given along with the values of the analytic approximations.
}
\begin{center}
\begin{tabular}{|c|c|c|c|c|c|c|c|c|}
\hline
$x/a$ 	& $y/a$ & $z_{\rm Br}$ 	& Name 		& Energy~[meV]	& $U(x,y)$ 	&$s$~[e\AA{}]	&$s(x,y)$&$q$~[e]\\
 	& 	& [\AA{}] 	&  		& from DFT	& [meV] 	&from DFT	&[e\AA{}]&from DFT\\ \hline
0	& 0	& 6.7		&top		& 257		& 508 		&0.68		&1.10	 &-0.19\\ \hline
0	& 0.25	& 6.6		&		& 189		& 254		&0.56		&0.70	 &-0.17\\ \hline
0.25	& 0.25	& 6.5		&quarter 	& 137		& 137		&0.45		&0.46	 &-0.14\\ \hline
0	& 0.5	& 6.4		&bridge		& 0		& 0 		&0.30		&0.30	 &-0.11\\ \hline
0.25	& 0.5	& 6.3		&		& 21		& 21		&0.22		&0.22	 &-0.09\\ \hline
0.5	& 0.5	& 6.2		&hollow		& 34		& 42		&0.14		&0.14	 &-0.07\\ \hline
\end{tabular}
\end{center}
\label{tab:min}
\end{table}

\begin{table}
\caption{
Summary of results for the Br coverage dependence study.
}
\begin{center}
\begin{tabular}{|c|c|c|c|c|}
\hline
$\Theta$ 	&$a_{\rm Br}$	&$\Phi/N+\mathcal{B}$~[meV]	& $s$~[e\AA{}]	&$q$~[e]	\\
		&[\AA{}]	&from DFT			&from DFT	&from DFT	\\ \hline
$1/9$	 	&8.91		&-2568				&0.31		&-0.12		\\ \hline
$2/9$	 	&6.3		&-2557				&0.32		&-0.12		\\ \hline
$1/4$	 	&5.94		&-2532				&0.30		&-0.12		\\ \hline
$1/2$	 	&4.2		&-2522				&0.27		&-0.13		\\ \hline
$1$	 	&2.97		&-1465				&--		&--		\\ \hline
$2$	 	&2.1		&5590				&--		&--		\\ \hline

\end{tabular}
\end{center}
\label{tab:cov}
\end{table}

\clearpage

\begin{figure}[ht]
\begin{center}
\includegraphics[width=0.75\columnwidth]{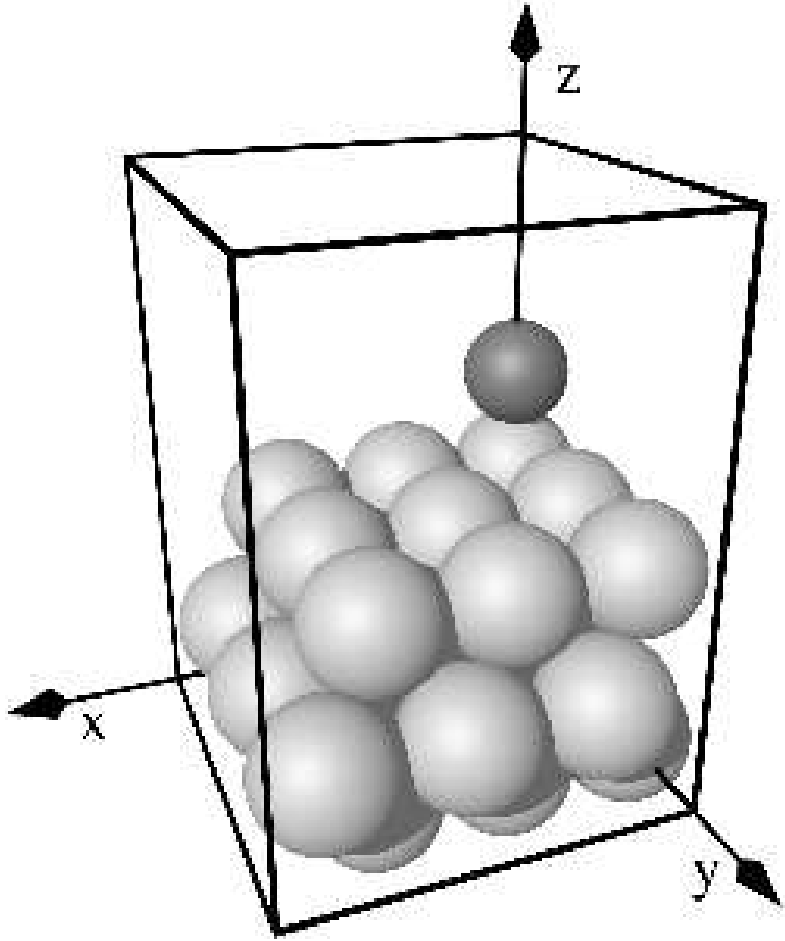}
\caption[]
{
A typical $3 \times 3$ supercell containing three metal layers,
(light gray spheres) and one Br atom (dark gray sphere). 
The cell is $3 ( 2.97 ) \times 3 ( 2.97 ) \times 12.6$ \AA{} in size.
}
\end{center}
\label{fig:supercell}
\end{figure}

\clearpage

\begin{figure}[ht]
\begin{center}
\includegraphics[width=0.75\columnwidth]{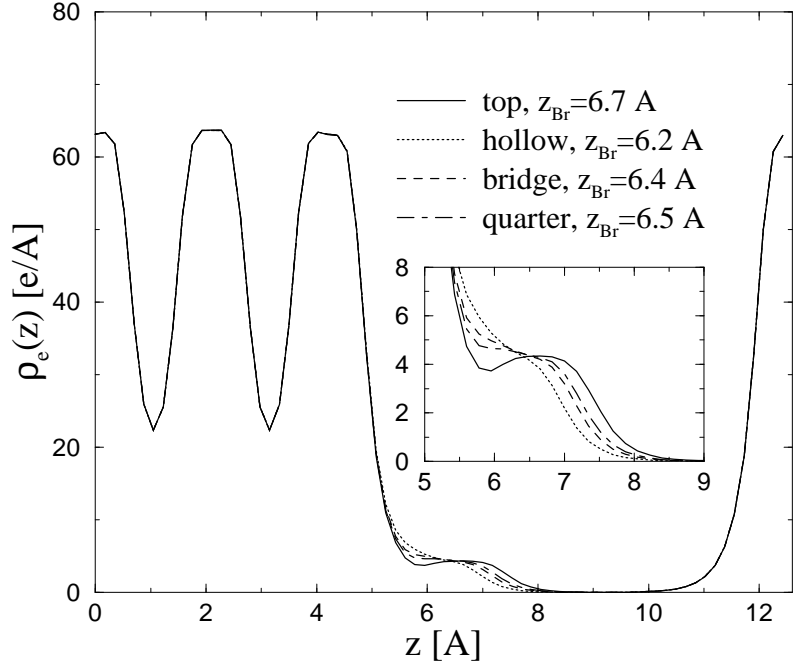}
\caption[]{
Electronic charge density, $\rho_{\rm e} ( z )$, 
for the lowest energy configurations for several binding sites, summarized in Table~\ref{tab:min}. 
Few meaningful differences are seen outside of the adlayer region, even in a magnified view of the surface region (inset).
$\Theta=1/9$.
}
\end{center}
\label{fig:chgzlat}
\end{figure}

\clearpage

\begin{figure}[ht]
\begin{center}
\includegraphics[width=0.75\columnwidth]{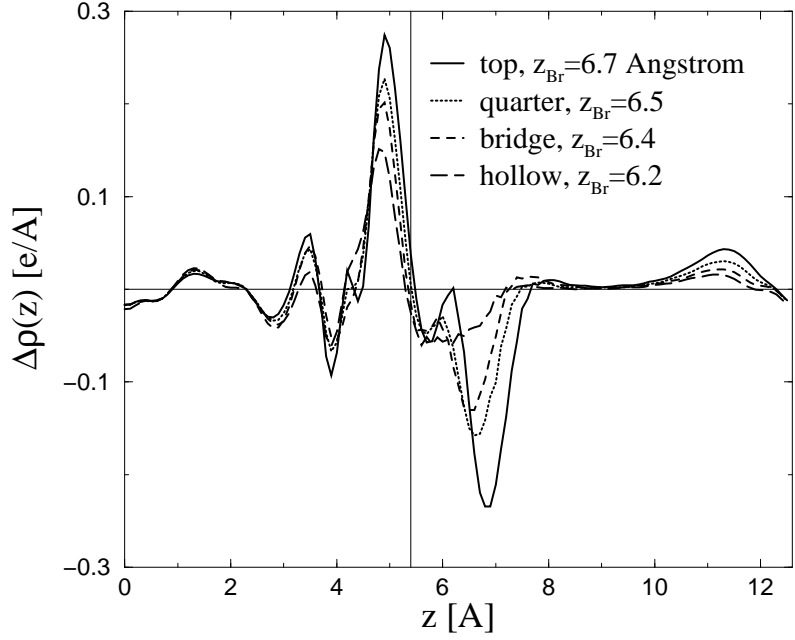}
\caption[]{
The charge transfer function, $\Delta \rho ( z )$, for the lowest
energy configurations for several binding sites, summarized in Table \ref{tab:min}. 
The straight horizontal line is used to indicate $\Delta \rho (z ) = 0$, 
and the vertical line indicates $z = 5.4$ \AA{}. 
The charge polarization in the surface region is easily observed.
$\Theta=1/9$.
} 
\end{center}
\label{fig:ctzmin}
\end{figure}

\clearpage

\begin{figure}[ht]
\begin{center}
\includegraphics[width=0.75\columnwidth]{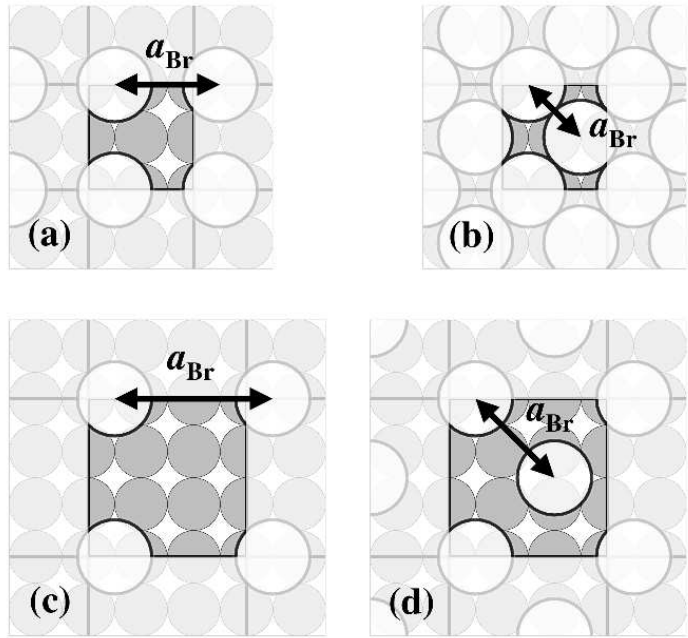}
\caption[]{
Four adlayer supercells. 
Small gray circles represent the top layer of Au(100) atoms, and the larger white circles represent adsorbed Br. 
Each cell has been repeated in each lateral direction, 
and the repeated cells are shown by lighter shading. 
(a) $2 \times 2$ cell with $\Theta = 1 / 4$, 
(b) $2 \times 2$ cell with $\Theta = 1 / 2$, 
(c) $3 \times 3$ cell with $\Theta = 1 / 9$, 
(d) $3 \times 3$ cell with $\Theta = 2 / 9$. 
The arrows indicate the adlayer lattice spacing, $a_{\rm Br}$, for each cell.
}
\end{center}
\label{fig:cells}
\end{figure}

\clearpage

\begin{figure}[ht]
\begin{center}
\includegraphics[width=0.75\columnwidth]{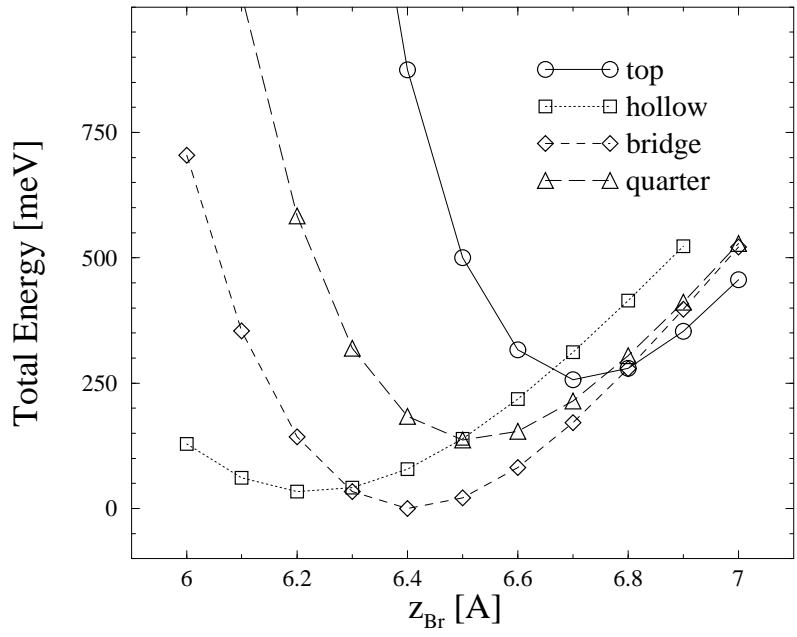}
\caption[]{
Total energy from DFT calculations vs $z_{\rm Br}$ for several lateral positions. 
See Table~\ref{tab:min} for more information.
$\Theta=1/9$.
} 
\end{center}
\label{fig:enerz}
\end{figure}

\clearpage

\begin{figure}[ht]
\begin{center}
\includegraphics[width=0.75\columnwidth]{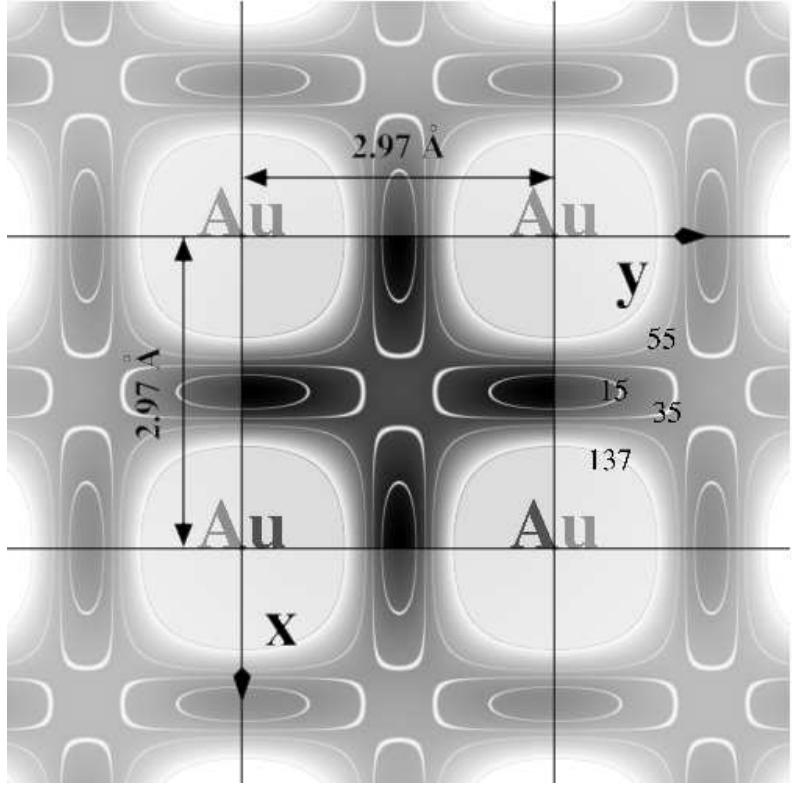}
\caption[]{
Sinusoidal corrugation potential, $U(x,y)$. 
The grid lines indicate the boundaries between surface unit cells, and the underlying Au(100) atoms are labeled. 
The grayscale image in the central square shows $U ( x, y )$, 
where black indicates the lowest value, 0~meV, and white indicates $U ( x, y ) = 137$~meV. 
The gray shading is truncated above $U ( x, y ) = 137$ meV, 
and the contour lines indicate $U ( x, y ) = 15$, $35$, $55$, and $137$~meV as labeled in the plot.
}
\end{center}
\label{fig:corr}
\end{figure}

\clearpage

\begin{figure}[ht]
\begin{center}
\includegraphics[width=0.75\columnwidth]{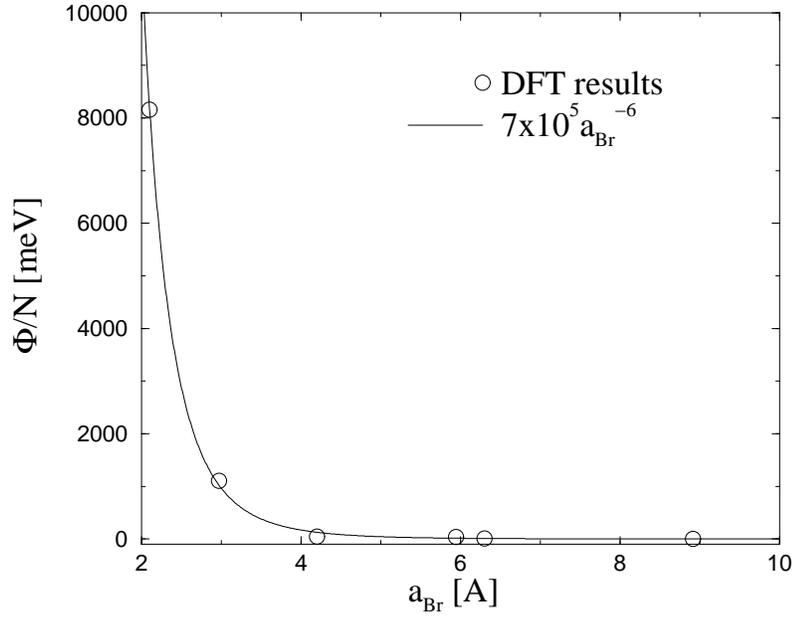}
\caption[]{
Lateral interactions determined by DFT calculations
along with an approximate analytic form.
}
\end{center}
\label{fig:latinter}
\end{figure}

\clearpage

\begin{figure}[ht]
\begin{center}
\includegraphics[width=0.75\columnwidth]{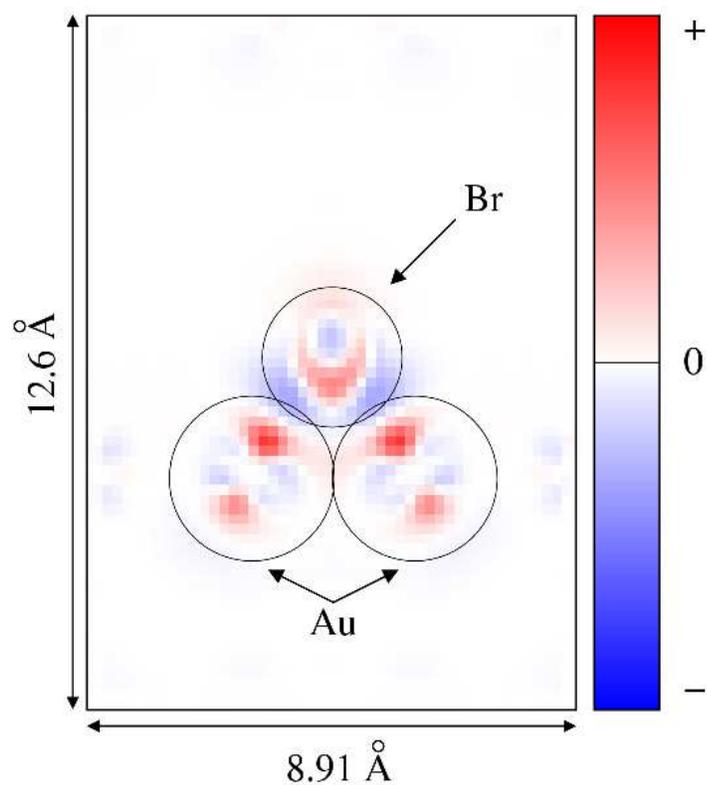}
\caption[]{
A two-dimensional slice through the three-dimensional charge transfer function, 
sliced through the two top sites to show the polar surface bonds associated with bridge site binding. 
The charge transfer function is for a $3 \times 3$ cell containing a single Br adsorbed at the bridge site. 
The circles indicate the adsorbed Br atom and the two Au atoms nearest to the Br, as labeled in the figure. 
The color scale indicates the local electron density,
as indicated in the legend to the right.
$\Theta=1/9$.
}
\end{center}
\label{fig:slice}
\end{figure}

\clearpage

\begin{figure}[ht]
\begin{center}
\includegraphics[width=0.75\columnwidth]{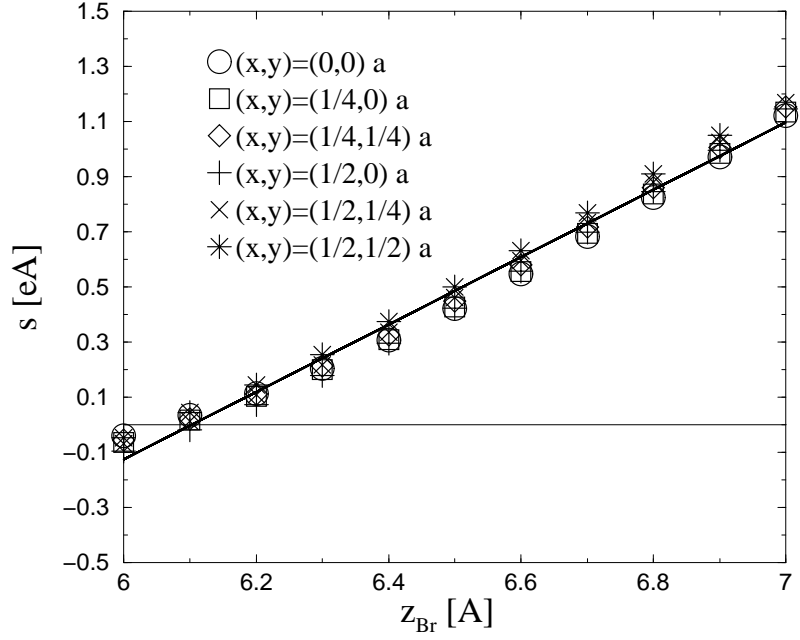}
\caption[]{
The surface dipole moment, $s$, vs.\ $z_{\rm Br}$ for all considered lateral configurations with $\Theta=1/9$.
It is quite clear, that even though the charge transfer function has a strong dependence on the lateral location, 
the surface dipole moment is essentially independent of lateral location and has a nearly linear dependence on $z_{\rm Br}$. 
The horizontal line indicates $s = 0$, and the heavy line indicates a linear fit to all data points.
}
\end{center}
\label{fig:diplat}
\end{figure}

\clearpage

\begin{figure}[ht]
\begin{center}
\includegraphics[width=0.9\textwidth]{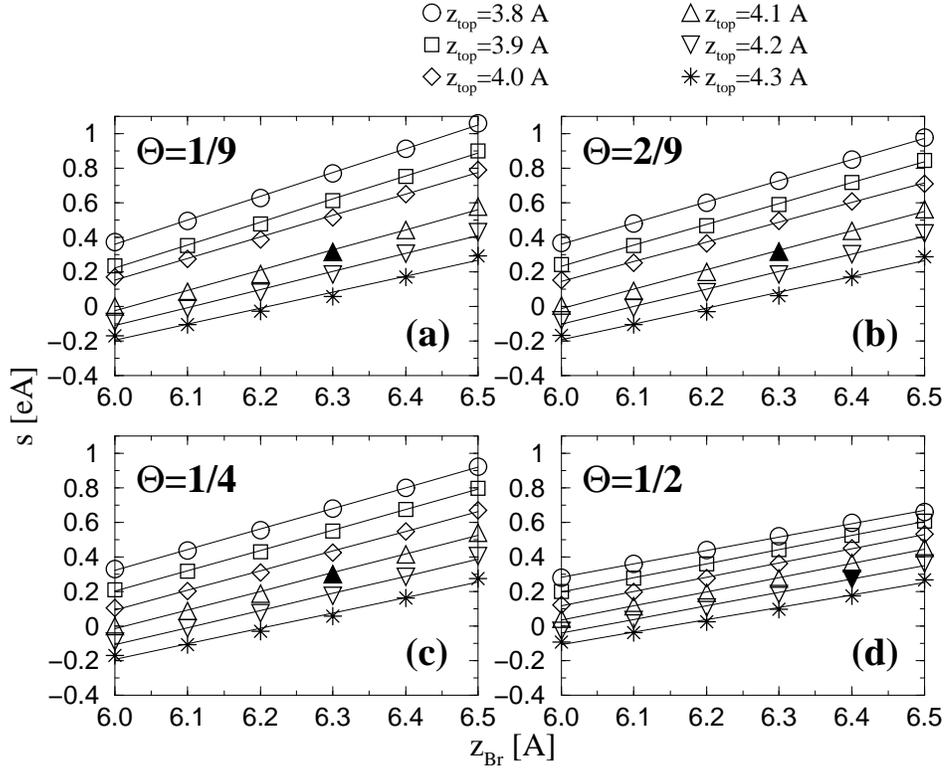}
\caption[]{
The surface dipole moment for four coverages, 
$\Theta =$ $1 / 9$ (a), $2 / 9$ (b), $1 / 4$ (c), and $1 / 2$ (d). 
The shaded symbols represent the configuration corresponding to the minimum energy for each coverage.
}
\end{center}
\label{fig:alldipmoms}
\end{figure}

\clearpage

\begin{figure}[ht]
\begin{center}
\includegraphics[width=0.75\columnwidth]{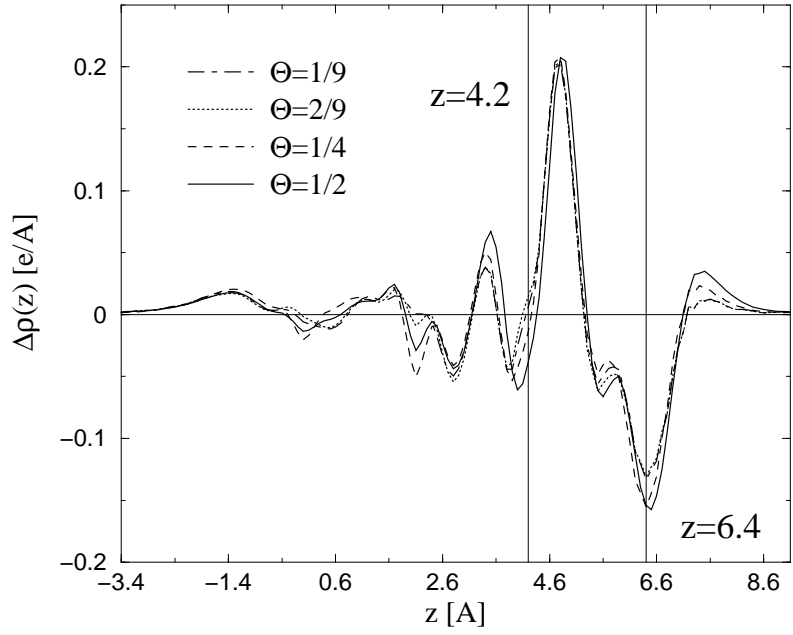}
\caption[]{
The charge transfer function, $\Delta \rho ( z )$. 
The straight horizontal line is used to indicate $\Delta \rho ( z ) = 0$, 
and the two vertical lines indicate $z_{\rm top} = 4.2$~\AA{} and $z_{\rm Br} = 6.4$~\AA{}.
}
\end{center}
\label{fig:chgzdiff}
\end{figure}

\clearpage

\begin{figure}[ht]
\begin{center}
\includegraphics[width=0.75\columnwidth]{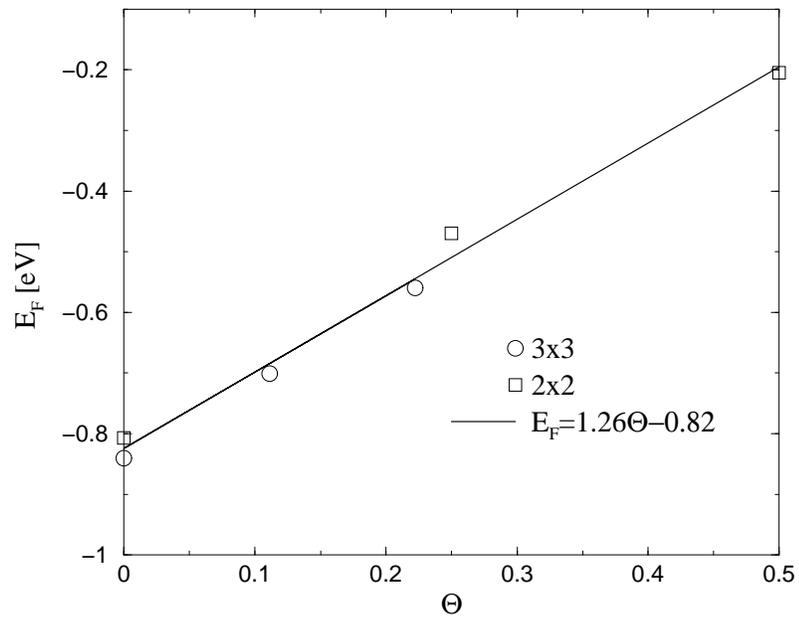}
\caption[]{
The Fermi energy vs. coverage for $2\times 2$ and $3 \times 3$ cells.
The linear dependence indicates coverage independence of the charge distribution.
}
\end{center}
\label{fig:efermi}
\end{figure}

\clearpage

\begin{figure}[ht]
\begin{center}
\includegraphics[width=0.75\columnwidth]{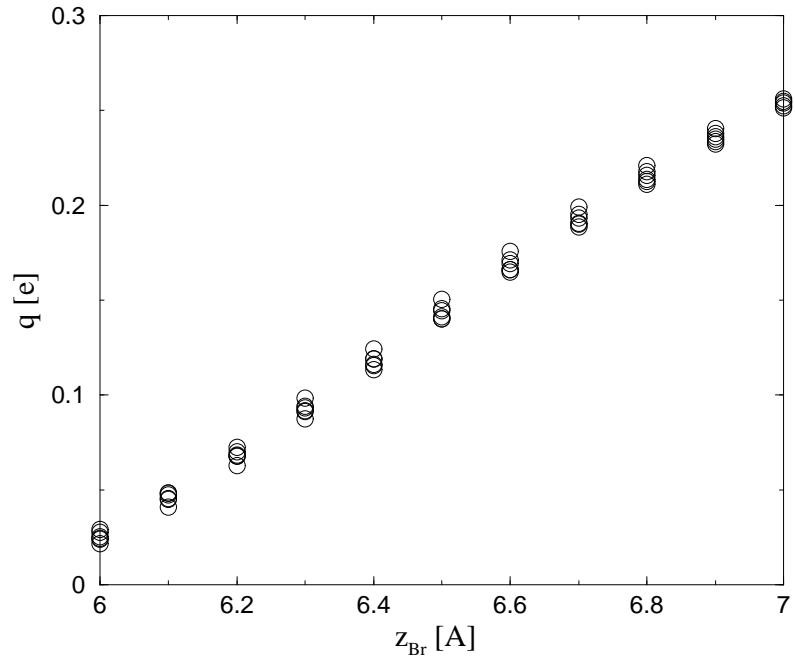}
\caption[]{
The residence charge vs. $z_{\rm br}$ for all considered lateral configurations with $\Theta=1/9$.
The linear dependence indicates that a charge/image charge description of the adlayer is not correct.
}
\end{center}
\label{fig:qz}
\end{figure}

\clearpage

\begin{figure}[ht]
\begin{center}
\includegraphics[width=0.75\columnwidth]{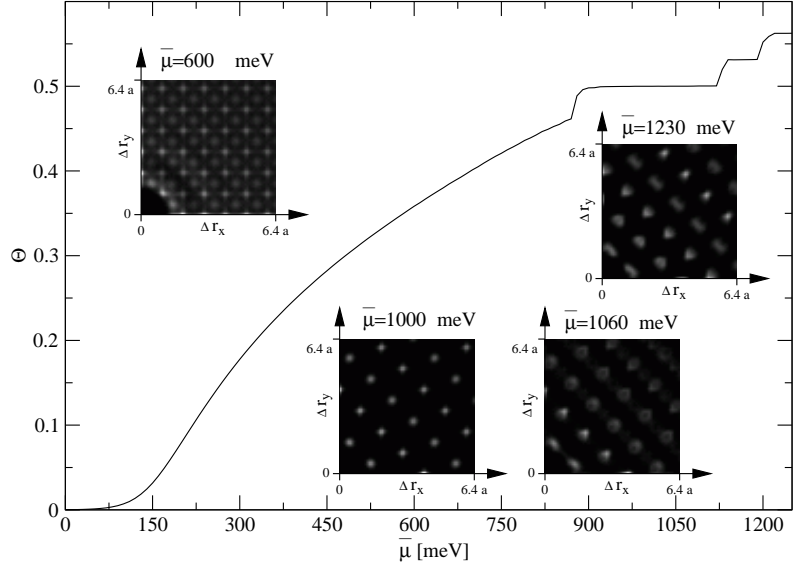}
\caption[]{
The results of the off-lattice Monte Carlo simulations for an $L=32$ surface.
Four different phases of the Br adlayer are seen in the simulations.
The three ordered phases are seen as coverage plateaus.
The insets show the short-range two point correlation function for each of these phases,
shown in grayscale,
where white represents places of highly coordinated atomic positions.
An atom always sits at the origin of each of the insets,
and ``a'' indicates the surface Au(100) lattice spacing, $a$.
}
\end{center}
\label{fig:offlattcov}
\end{figure}

\clearpage

\begin{figure}[ht]
\begin{center}
\includegraphics[width=0.75\columnwidth]{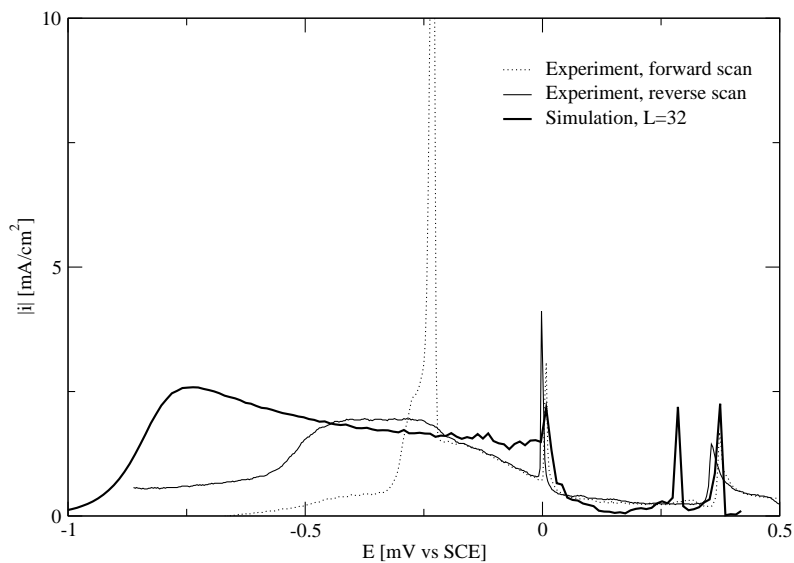}
\caption[]{
The results of the off-lattice Monte Carlo simulations compared to the experimental cyclic voltammetry results.
The experimental results are taken from Ref.~\cite{WAND96} with permission from the authors.
Note that peak heights are very difficult to calculate by Monte Carlo methods,
and the peak positions are far more important.
}
\end{center}
\label{fig:offlatt}
\end{figure}

\end{document}